\documentstyle[12pt]{article}
\begin{document}
\rightline{{\bf CWRU-P11-95, CITA-95-18}}
\rightline{{\bf astro-ph/9509115 }}
\rightline{September 1995}
\baselineskip=16pt
\vskip 0.2in

\newcommand{\beq}{\begin{equation}}
\newcommand{\eeq}{\end{equation}}
\newcommand{\feh}{\hbox{$[{\rm Fe}/{\rm H}]$}}
\newcommand{\mvrr}{\hbox {$\rm M_v(RR)$}}
\newcommand{\dvtwo}{\Delta {\rm V}}
\newcommand{\dv}{\hbox {$\Delta \rm V^{TO}_{HB}$}}
\newcommand{\ea}{{\it et al.}}
\newcommand{\la}{\mathrel{\hbox{\rlap{\hbox{\lower4pt\hbox{$\sim$}}}\hbox{$<$}}}}
\newcommand{\ga}{\mathrel{\hbox{\rlap{\hbox{\lower4pt\hbox{$\sim$}}}\hbox{$>$}}}}

\begin{center}
{\bf\large
A Lower Limit on the Age of the Universe}
\end{center}
\vskip 0.15in
\begin{center}
Brian Chaboyer
\\{\small\it CITA, \\60 St. George St., Toronto, ON, Canada M5S 1A7}
\vskip 0.13in
Peter J. Kernan and Lawrence M. Krauss
\footnote{Also Dept of Astronomy.}
\\
{\small\it Department of Physics\\
Case Western Reserve University\\ 10900 Euclid Ave., Cleveland, OH
44106-7079}
\vskip 0.13in
Pierre Demarque
\\
{\small\it Department of Astronomy, Yale University, New Haven CT
06520-8101}
\end{center}
\vskip 0.1in
\centerline{{\bf Abstract}}
\noindent
We report the results of a detailed numerical study
designed to estimate both the absolute age and the
uncertainty in age (with confidence limits) of the oldest
globular clusters.  Such an estimate is essential
if a comparison with the Hubble age of the universe is to be made to
determine the consistency, or lack thereof, of various cosmological
models. Utilizing estimates of the uncertainty range (and distribution)
in the input parameters of stellar evolution codes we produced 1000
Monte Carlo realizations of stellar isochrones, with which we
could fit the ages of the 18 oldest globular clusters.  Incorporating
the observational uncertainties in the measured color-magnitude
diagrams for these systems and the predicted isochrones, we derived a
probability distribution for the mean age of these systems.   The
one-sided 95$\%$ C.L.~ lower bound for this distribution occurs at an
age of 12.07 Gyr. This puts interesting constraints on cosmology
which we discuss. Further details, including a description of the
distributions, covariance matrices, dependence upon individual input
parameters, etc.\  will appear in a future article.

\newpage
\baselineskip=20pt

The apparent dichotomy between the upper bound on the age of the
Universe obtained from the Hubble Constant  and the lower bound
obtained by dating the oldest globular clusters in our galactic halo
represents one of the most significant potential conflicts in modern
observational cosmology.  For Hubble constant $H_0 =100 \ h \
{\rm km\, sec^{-1}Mpc}^{-1}$, a flat, matter dominated universe has an
age given by
\begin{equation}
\tau_{Hubble} = 2/3 \ H_0^{-1} = 6.6 h^{-1} Gyr
\end{equation}

If the universe is open, the factor of 2/3 is changed to unity, but
even in this case if the age of the oldest globular clusters in our
galaxy is really 16 Gyr, as various best fits estimates recently
suggest \cite{dem,vict}, then this will be inconsistent with the Hubble
age if
$h>.6$.  Consistency with the flat matter dominated universe estimate is
impossible if $h >.4$.   Since several recent estimates of the Hubble
constant based on Type 1a supernova, and Hubble Space Telescope
observations of Cepheid variables in the Virgo cluster both
tend to suggest $h
> .65 $ \cite{Riess,Freedman}, this has been one factor which has led
various groups to argue once again for the need for a cosmological
constant (i.e. \cite{krauss}).

Since the Hubble estimate is unambiguous for a fixed Hubble
constant the crucial uncertainty in this comparison resides in the
globular cluster (GC) ages estimates themselves.  Rough arguments have
been made that changes in various input parameters in the stellar
evolution codes designed to derive globular cluster isochrones, or in
the RR Lyrae distance estimator used to determine absolute
magnitudes for GC stars, might change age estimates by 10-20 $\%$ (i.e.
\cite{chab}).  However, no systematic study has yet been undertaken to
realistically estimate the cumulative effect of all existing
observational and theoretical uncertainties in the GC
age analysis.  This is the purpose of the present work.

One of the reasons we believe such an analysis had not yet been
carried out is that it is numerically intensive.  Each
run of a stellar evolution code for a single mass point takes
3-5 minutes on the fastest commercially available workstations.
Nine different mass points
at three different metallicity values must be run to produce each
set of isochrones.
If one then runs, say, 1000 different isochrone sets to explore the
different parameter ranges available, this requires over 8 weeks of
continuous processing time.

This is long, but not prohibitive, so because of the importance of this
issue, we developed the Monte Carlo algorithms necessary for
the task.  This involved first examining the measurements of input
parameters in the stellar evolution code to determine their best fit
values, and also their uncertainties along with the appropriate
distributions to use in the Monte Carlo.  Then the stellar
evolution code
\cite{guenther} and isochrone generation code were rewritten to allow
sequential input of parameters chosen from these distributions, and
output of the necessary color-magnitude (CM) diagram observables.
Finally, we derived a fitting program to compare the predictions to
the data.   Since the numerically intensive part of this
procedure involves the Monte Carlo generation of isochrones, by
incorporating the chief observational CM uncertainty afterwards our
results can quickly be refined as this uncertainty is
refined.

In this article we briefly describe the general features of our
analysis and present our main result, the derived probability
distribution for the age of the oldest galactic globular clusters.
This should be the result which is of broadest and most immediate
general interest. In a future work we shall describe the details of the
analysis, and present our results for covariance matrices, correlation
functions, dependence on individual parameters, and observational
uncertainties.  These results will be of more use to researchers in the
field of stellar evolution and globular cluster dating.

\vskip 0.1in
\noindent{\bf Monte Carlo Analysis Inputs:  General Features}

 We have focussed on what we believe are the chief input uncertainties
in the derivation of stellar evolution isochrones.  These include: pp
and CNO chain nuclear reaction rates,
stellar opacity uncertainties, uncertainties in the treatment of
convection and diffusion, helium abundance uncertainties,
and uncertainties in the abundance of the $\alpha$-capture elements
(O, Mg, Si, S, and Ca).  Our stellar evolution code was revised to
allow batch running with sequential input of these parameters chosen
from underlying probability distributions.  We did not include the
equation of state among our Monte Carlo variables as it is now well
understood in metal-poor main sequence stars.  It has recently been
shown that the detailed equation of state by Rogers \cite{rogers}
gives very similar globular cluster age estimates to those obtained
using the Debye-H\"{u}ckel correction \cite{chabeos}, which was used in
this study.

We describe briefly our parameter choices and distributions below:
We included uncertainties for the 3 most important reactions
in  the pp chain. For $p + p\to ^2$H$ + e^+ + \nu$ we used the analysis
of Kamionkowski and Bahcall (\cite{KB}), except that where they
took theoretical errors as gaussianly distributed, we decided that
a uniform distribution better represented the state of our
(lack of) knowledge. There are two sources of theoretical error,
one from the uncertainty in the particle wave-functions, another
from meson exchange (\cite{barg}). We use a relative modification
to this reaction of $1 \pm .002 \ ^{+.0014}_{-.0009} \ ^{+.02}_{.012}$,
where the 2nd term is 1-sigma gaussian, and the 3rd and 4th are
top hat distributions. For 2 other {\em pp} chain reactions,
$^3$He + $^3$He $\to$ $^4$He + p + p, and
$^3$He + $^4$He $\to$ $^7$Be + $\gamma$,
the uncertainties included in this study were taken from
Bahcall and Pinsonneault (\cite{BP}, Table I).
The CNO nuclear reaction rates
and their uncertainties are from Bahcall (\cite{bahcall}; Table 3.4).

The
stellar opacities are broken into high temperature and low temperature
regimes.  For the high temperature regime ($T>10^4\,$K), the OPAL
opacities \cite{iglesias} are utilized.  The uncertainties in these
opacities were evaluated by a comparison to the LAOL opacities
\cite{huebner}.  In the temperature regime relevant for nuclear fusion
($T\ga 6\times 10^6\,$K), typically differences of 1\% were found,
with maximum differences of 3\%.  Thus, we elected to multiply the high
temperatures opacities by a Gaussian distribution, with a mean of 1,
and $\sigma = 0.01$.  The Kurucz opacities \cite{kuropac} were used
for the low-temperature regime.  Low temperature
opacities calculated by different groups can differ by a large amount
\cite{lowtopac}, though modern calculations appear to agree to within
$\sim 30\%$.  In addition, we intercompared Kurucz's calculations for
different element mixtures, finding maximum differences of 30\%.  For
these reasons, we elected to multiply the Kurucz opacities by a
number which was uniformly drawn from the range 0.7 -- 1.3.

The correct treatment of the convective regions of the star has been a
long standing problem in stellar astrophysics, which stems from our
poor understanding of highly compressible convection and its interaction
with radiation in the optically thin outer layers.  Most stellar
evolution codes use the mixing length approximation, which is a simple
model of convection where blobs of matter are supposed to rise or fall
adiabatically over some distance, and then instantaneously release their
heat \cite{bohm}.  The
uncertainties in this treatment of convection are parameterized into a
single non-dimensional quantity called the mixing length,
which is usually taken to be fixed during a
star's evolution.  Its value is typically chosen
by requiring that a solar model have the correct radius and
luminosity at the solar age, or by a comparison of theoretical
isochrones to observed CM diagrams.  A mixing length of around 1.8
appears to provide a reasonable match to the observations, but its
exact value depends on the input physics (opacities, model
atmospheres, etd.) used to construct the stellar models.
Modern solar models typically employ mixing lengths between
1.7 and 2.1 \cite{chabsun,chabsun2}.  To further explore this issue,
we have conducted tests whereby isochrones were constructed from
models with different mixing lengths and compared to a metal-poor
globular cluster CM diagram.
We found that changes in the mixing length of
0.3 could be ruled out as the isochrones no longer fit the data, if
all other input parameters were held constant.
Allowing for possible variation among globular clusters
and for the fact that a broader
range in mixing length might fit the data if other parameters are
allowed to vary at the same time,
we elected to use a rather broad Gaussian distribution
with a mean of 1.85 and $\sigma = 0.25$.

Whether or not to include the effects of element diffusion (whereby
helium tends to sink to the center of the star, while hydrogen is
rises to the surface) is a difficult question.  Physical models of
compressible plasmas suggest that diffusion should be occurring in
stars, and predict diffusion coefficients with a claimed accuracy of
about 30\%
\cite{michaud}. However, these models assume that no other mixing
process occurs within the radiative regions of a star, which may not
be true.  Helioseismology appears to suggest that diffusion is
occurring in the sun \cite{christian}, but the evidence is not
compelling
\cite{chabsun}.  Models of halo stars which incorporate diffusion
predict curvature in the Li-effective temperature plane which is not
observed \cite{chabli}, which suggests that some process is inhibiting
diffusion in halo stars.  Given these uncertainties, we have elected
to multiply the diffusion coefficients given by Michaud \& Proffitt
\cite{michaud} by a number uniformly drawn from the interval
0.3 -- 1.2.  The use of a flat distribution with a rather large range
reflects our conviction that the incorporation of diffusion within
stellar models is subject to large uncertainties.

The primordial helium abundance, relevant to old halo stars, and an
important input in our stellar codes, has taken on renewed interest as
a result of recent calculations of Big Bang Nucleosynthesis (BBN) light
element production
\cite{KK1,CST,KK3}.  In order to BBN estimates to agree with inferred
primordial abundances, significant systematic uncertainties must be
allowed for.  It has become clear that such uncertainties are the
dominant feature of the comparison between theory and observation.
We therefore utilize here a flat distribution
for the primordial helium mass fraction between 0.22 and 0.25, which
encompasses the range of recent estimates.

In order to calculate a stellar model, the abundance of the elements
heavier than helium (denoted by $Z$) must be specified.  Due to its
numerous spectral lines, it is relatively easy to determine the
abundance of Fe in globular cluster stars.  Unfortunately, the
abundances of the other heavy elements are more difficult to determine
and it has been common to assume that the other heavy elements are
present in the same proportion as they are in the Sun.  However, from
both theoretical arguments and observational evidence, it is clear that
the elements which are produced via $\alpha$-capture are enhanced in
abundance relative to their solar value.  It is relatively easy to
incorporate the effects of the enhancement of the
$\alpha$-elements on the stellar models by re-defining the
relationship between the model Z and the iron abundance
\cite{salaris}.  Oxygen is by far the most important of the $\alpha$-capture
elements, as it acounts for roughly half of all the heavy elements (by
number) present in the Sun.  For this reason, we concentrate on
observations of oxygen as representative of the
$\alpha$-capture elements.  The determination of oxygen abundances in
stars is extremely difficult, and subject to a number of systematic
uncertainties (e.g.~\cite{king}).  Recently, high quality
[O/Fe]\footnote{We use the common spectroscopic
notation, where the abundance of element $y$ relative to element $x$
is denote by $[y/x] \equiv \log (y/x)_{\rm star} - \log (y/x)_{\rm
sun}$.}  abundances for a number of halo stars have been obtained
\cite{nissen}, with the result that the mean abundance was found to
be $[{\rm O/Fe}] = 0.55\pm 0.05$, where the error is simply the
standard deviations of the measurements.  In addition to this error,
one must add in the possible systematic errors which have been
discussed by a number of authors \cite{king,tomkin,bessell}.  An
analysis of the literature has lead us to conclude that possible
systematic errors in the determination of $[\rm O/Fe]$ may be as large
as $\pm 0.2$ dex.  Thus, the abundance of the $\alpha$-elements was
taken to be $[\alpha/{\rm Fe}] = 0.55\pm 0.05\pm\,$(Gaussian) $\pm
0.2\,$(top-hat).

Finally, a color table must be used to convert our
theoretical luminosities and temperatures to observed magnitudes and
colors. The construction of an accurate color table requires the use
of theoretical model atmospheres, which are still subject to large
uncertainties.  We elected to take this uncertainty into account by
randomly choosing one of two totally independent color tables
\cite{green,kurcol}
with equal probability in constructing each isochrone set in
the Monte Carlo. These two tables span reasonably the
present range used to
transform from theoretical temperatures and luminosities to
observed colors and magnitudes.

\vskip 0.1in
\noindent{\bf The Fitting Procedure and the Probability Distribution
for Globular Cluster Ages}

Once a set of isochrones (which consists of isochrones of
three different metallicities for a set of 15 different ages
between 8 and 22 Gyr) is derived, the comparison with the observed
parameters of a specific set of globular clusters requires a fitting
procedure.  In order to minimize the large uncertainties in the
effective temperatures of the models \cite{renzini}, we have elected
to use the difference in magnitude between the main sequence
turn-off, and the horizontal branch (HB, in the RR Lyr instability
strip) as our age diagnostic.  This age determination technique is
commonly referred to as \dv ~and has been extensively used in the
astronomical literature (e.g.~\cite{carney}).  Our isochrone sets
provide us with the main-sequence turn-off luminosity as a function of
age and metallicity.  Due to the importance of convection in the
nuclear burning regions of HB stars, theoretical HB luminosities are
subject to large uncertainties, and so we have elected to combine our
theoretical main-sequence turn-off luminosities with an observed
relation for the luminosity of the HB (see below).  This results in a
grid of predicted \dv 's as a function of age and \feh, which
is then fit to an equation of the form
\beq
t_9 = \beta_0 + \beta_1\dvtwo + \beta_2\dvtwo^2 + \beta_3\feh +
\beta_4\feh^2 + \beta_5\dvtwo\feh,
\label{fit}
\eeq
where $t_9$ is the age in Gyr.  The observed values of \dv ~and \feh,
along with their corresponding errors, are input in
(\ref{fit}) to determine the age and its error for each GC in our
sample.

There is abundant evidence for a large age range within different GC
systems (e.g.~\cite{chaboyer,buonanno}) so one must take care to
select a sample which only includes old globular clusters.
Observational errors in the determination of the turn-off and
horizontal branch magnitudes lead to a $\sim 10-20\%$ error in the
derived age of any single cluster.  Thus, to minimize the
observational uncertainties, it is best to determine the mean age of a
number of GCs.  In light of the strong evidence for an
age-metallicity relationship (with metal-poor clusters being the
oldest), only metal-poor clusters were selected ($\feh \le -1.6$).
 From this list of metal-poor clusters, any cluster which has been
shown to be young using the difference in color between the
giant-branch and main sequence turn-off \cite{buonanno}, or which is
suspected of being young due to its unusually red horizontal branch for
its metallicity \cite{lee} was discarded.  From the sample of 43
GCs \cite{chaboyer} for which high quality observations
are available, 27 survived the metallicity cut, of which 10 were
discarded for being young, leaving a total of 17 globular clusters.
Our final sample contained the following clusters: NGC
1904, 2298, 5024, 5053, 5466, 5897, 6101, 6205, 6254, 6341, 6397,
6535, 6809, 7078, 7099, 7492, and Terzan 8.

One of the things we checked is the the inferred dispersion in the age
of the 17 globular clusters is not larger than that which we expect
based on the observational uncertainties.  Using the uncertainties
in the individual ages determined from uncertainties in the observed
turnoff magnitude and metallicity, we examined the dispersion about
the mean age for the 17 clusters using a $\chi^2$ test.  We found
a reduced $\chi^2$ of 0.55 per degree of freedom, indicating both
no evidence for any intrinsic dispersion in age for our sample,
 and that
the quoted observational uncertainties for each cluster may be too generous.
In any case, given the
quoted accuracy, it is certainly consistent to assign a single mean age for
the sample.

One chief observational uncertainty common to all globular
clusters is retained until the end of the analysis.
This is the  $M_v$ determination
for RR-Lyrae variables.
In order to determine \dv ~as a function of age and metallicity, we
combine our theoretical turn-off magnitude with an observationally
based estimate for the absolute magnitude of the HB, in the RR Lyr
instability strip (hereafter referred to as \mvrr).  There are a
number of independent, observationally based techniques which can be
used to derive \mvrr.  In general, it has been found that the absolute
magnitude of the RR Lyr stars can be represented by an equation of the
form
\beq
\mvrr = \mu\, \feh + \gamma,
\label{rrlyr}
\eeq
where $\mu$ is the slope with metallicity and $\gamma$ is the
zero-point.  Note that \mvrr ~is independent of age (at least, with
systems greater than 8 Gyr old).  Recent estimates for the slope with
metallicity vary from 0.15 to 0.30 \cite{carney,chaboyer}.  Fortunately, since
we are determining the mean age of 17 globular clusters in the
restricted metallicity range $-2.41 \le \feh \le -1.60$, the
uncertainties in the slope has only a small effect on our age
estimate.  Tests which we conducted indicated that the
maximum difference
in our mean age was only 0.5\% when the slope was varied between 0.15
and 0.30.  The most recent work suggests that $\mu = 0.20$ is likely
to be correct \cite{jones,skillen} which is the value we have used in
this study.  Uncertainties in the zero-point, $\gamma$ in
eq.~\ref{rrlyr} have a large impact on our derived age estimates.  For
this reason, we have spent considerable time reviewing recent
observational estimates of the zero-point.  These estimates for the
zero-point are usually given as a \mvrr ~value at a specific
metallicity.  As the globular clusters in our sample have a median
metallicity of $\feh = -1.82$ and a mean metallicity of $\feh = -1.93$
we have elected to transform the various zero-point estimates to
$\feh =-1.90$ using a slope of $\mu = 0.20\pm 0.04$.

Layden Hanson \& Hawley \cite{layden} have recently used the
statistical parallax technique to determine $\mvrr = 0.68\pm 0.12$ in
field halo RR Lyr stars.  Walker \cite{walker} measured the apparent
magnitude of LMC RR Lyr stars, and assumed an LMC distance modulus of
18.5 to infer $\mvrr = 0.44\pm 0.10$.  This choice for the distance
modulus of the LMC was based on the Cepheid distance, main sequence
fitting and the SN1987A ring distance to the LMC.  This last method is
a purely geometrical method, and should be the most reliable.
However, the SN1987A distance to the LMC has recently been revised to
18.37 \cite{gould}, implying $\mvrr = 0.57\pm 0.10$.  Main sequence
fitting of globular cluster CM diagrams to local halo stars with well
determined parallaxes can be used to determine the distance to
globular clusters, and hence, \mvrr.  Unfortunately there is only
one relatively metal-rich sub-dwarf which has a well determined
parallax.  Application of this technique to the globular cluster M5
yields $\mvrr = 0.76\pm 0.12$ \cite{carney}.  The only direct
determination of \mvrr ~in a metal-poor globular cluster is by Storm,
Carney \& Latham \cite{storm}.  They used a Baade-Wesselink/infrared
flux analysis to determine $\mvrr = 0.52\pm 0.26$.  Although the error
is large due to possible systematic uncertainties, it does suggest
that the RR Lyr stars in metal-poor globular clusters are somewhat
brighter than those found in the field \cite{layden}, or in metal-rich
globular clusters \cite{carney}.  In light of the above estimates, we
have elected to use $\mvrr = 0.60\pm 0.08$ (corresponding to $\gamma =
0.98\pm 0.08$).  This central value was chosen by a straight average
of the 4 published \mvrr ~estimates referenced above.  The error bar
was chosen to ensure that the $1\,\sigma$ range would include the
central value obtained for \mvrr ~in a metal-poor globular cluster,
and the $2\,\sigma$ range (0.44 -- 0.76) would encompass all of the
estimates quoted above.  Our choice is further supported by a
study which just appeared comparing four different distance estimators,
including kinematic, RR Lyr, Cepheid and Type II Supernovae
for consistency \cite{huterer}, and found that this appeared to
require a range for \mvrr ~similar to the one we chose.

Figure 1 displays the ensemble of age estimates from our Monte Carlo,
for different values of \mvrr.  It was generated as follows:
For each of the sets of
isochrones and given a value for \mvrr ~ we determine a mean age
and $1\sigma$ uncertainty in the mean. In the figure we show these
values for each isochrone set, for 3 values of \mvrr : the mean, .60,
and the endpoints of our 95\% C.L.~ range, .44 and .76.
In order to obtain the final histograms displayed as figs 2 a-b, we follow
an analogous procedure, but this time allow \mvrr ~to be a random
variable, and sample the sets of isochrones with replacement 12,000
times. For each sample, rather than the mean age, we record  a random
age drawn from a gaussian distribution with the mean age and variance
for that isochrone set at that \mvrr . Then the data are sorted and
binned to produce the figures:
2(a) the full distribution for
the assumed Gaussian spread in \mvrr ~of $\pm .08$, and
2(b) the distribution
for fixed value of \mvrr ~of 0.6.   In this way the effect of
the uncertainty in \mvrr ~can be explicitly examined.

\vskip 0.1in
\noindent{\bf Conclusions}

Our results indicate that at the one sided 95$\%$ confidence level
(determined by requiring 95$\%$ of the determined ages to fall above
this value) a lower limit of approximately 12.1 Gyr can be placed on the
mean value of these 18 Globular clusters.  (The symmetric 95$\%$
range of ages about the mean value of 14.56 Gyr is 11.6-18.1 Gyr.)  Note
that the distribution deviates somewhat from Gaussian, as one might
expect.  In particular, at the lower age limits the rise is steeper
than Gaussian, reflecting the fact that essentially all models give an
age in excess of 10 Gyr, while the tail for larger ages is larger than
gaussian.  The explicit effect of the largest single
common observational uncertainty, that in \mvrr , increases the net
width of the distribution by approximately $\pm 0.6$ Gyr
(i.e. $\approx \pm 5\% $).
Note that simply varying \mvrr over its full $2\sigma$ range,
keeping all other parameters fixed,
would produce a $ \pm 16\% $ change in GC ages estimates. For
comparison, the
next most significant input parameter uncertainties in this same
sense are [$\alpha$/Fe] ($\pm 7 \% $ effect), mixing length
($\pm 5\%$ effect), and diffusion, $^{14}$Np reaction rate, and
primordial Helium abundance, each of which would affect age estimates
at the $\pm3\%$ level if allowed to vary over its entire
range, keeping all other parameters fixed.

We believe that this result can now be used with some confidence to
compare to cosmological age estimates.   Of course, in addition to the
age determined here one must add some estimate for the time it took
our galactic stellar halo to form from the initial density
perturbations present during the Big Bang expansion. Estimates for this
formation time vary from 0.1 -- 2 Gyr.  To be conservative, we choose the
lower value.  In this case, we find that the age of
globular clusters in our galaxy is inconsistent with a flat, matter
dominated universe unless $h < 0.54$, and for a nearly empty, matter
dominated universe unless $h < 0.80$.  If the value of h is definitely
determined to be larger than either of these values, some
modification, such as the addition of a cosmological constant, would
seem to be required.

\vskip 0.3in
\noindent{We acknowledge computing support from the Ohio Supercomputer
Center.
LMK thanks CITA, the Aspen Center for Physics,
and the Institute for Nuclear and Particle Astrophysics, while PK thanks
CITA and the Aspen Center for Physics, and BC thanks the
Physics Dept at Case Western Reserve University, for their hospitality
during various stages of this work.
LMK's research is supported in part by the DOE.}

\newpage
\noindent{\bf \Large Figure Captions}

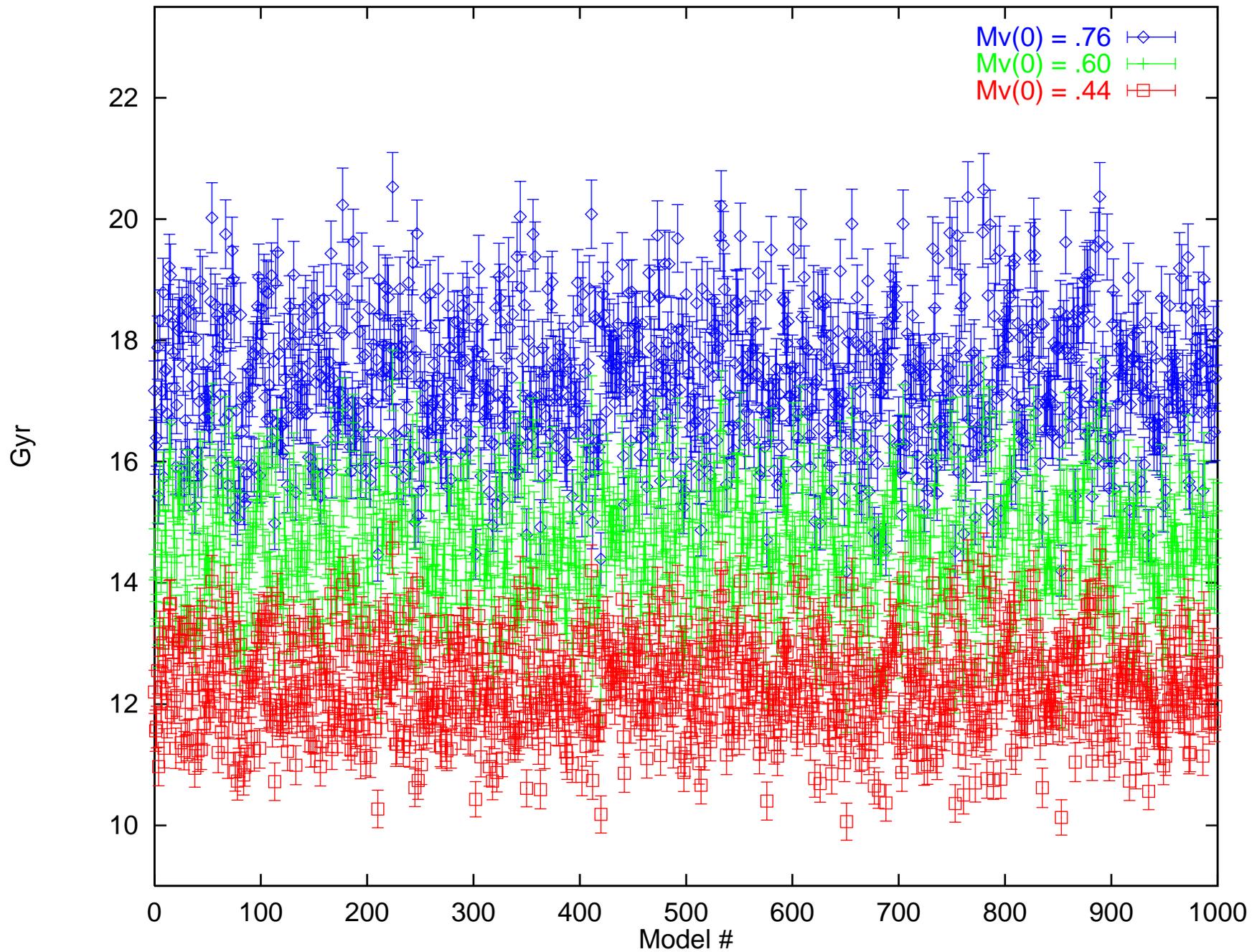
\noindent Figure 1: The suite of Models generated by the
Monte Carlo Procedure. For each Model the $1-\sigma$ age
range is plotted for 3 choices of \mvrr. The red, green
and blue data points show the age variation for \mvrr\ =
0.44, 0.60 and 0.76 respectively.

\noindent Figure 2: The histograms exhibit the relative numbers
of realizations of mean globular cluster ages
drawn randomly from the Monte Carlo data set (with uncertainties
on individual age estimates taken to be Gaussian) using
a value for \mvrr ~ (the absolute magnitude of the RR Lyrae Variables)
chosen from one of two different distributions:
a) a Gaussian, \mvrr\ = $0.6\pm .08$,  b) a Delta function,
\mvrr\ = 0.6.
The dashed line is a Gaussian
approximation to the actual distribution. The mean and standard
deviation of the Gaussians are shown in the Legend. The horizontal
arrows show the 1 and 2-$\sigma$ ranges in age based upon
the Gaussian approximation. The one-sided 95\%
Confidence Limit for a lower bound on the age of the Universe
is displayed as an arrow extending to the right from a vertical bar.
This is calculated directly from the generated distribution.

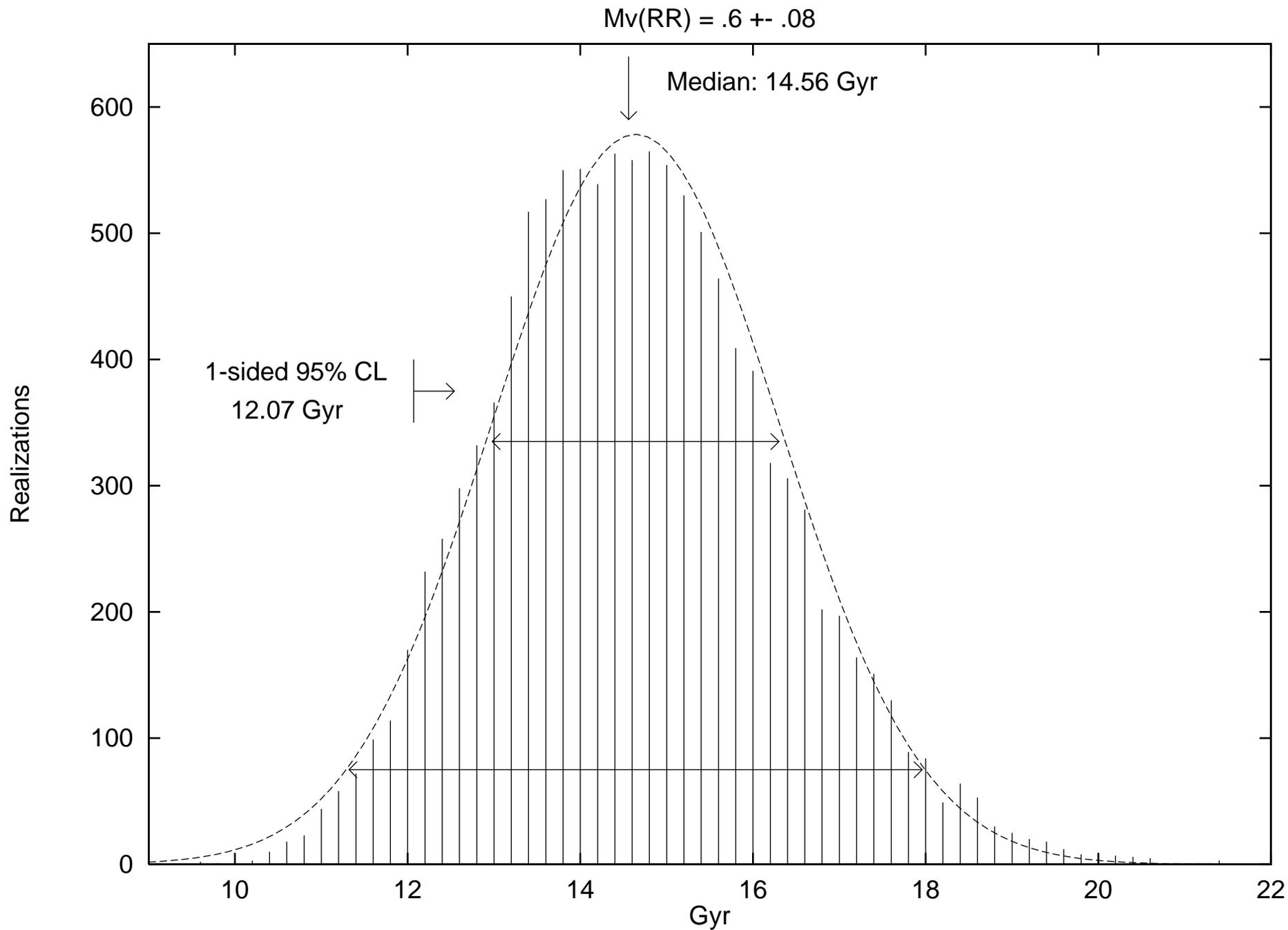
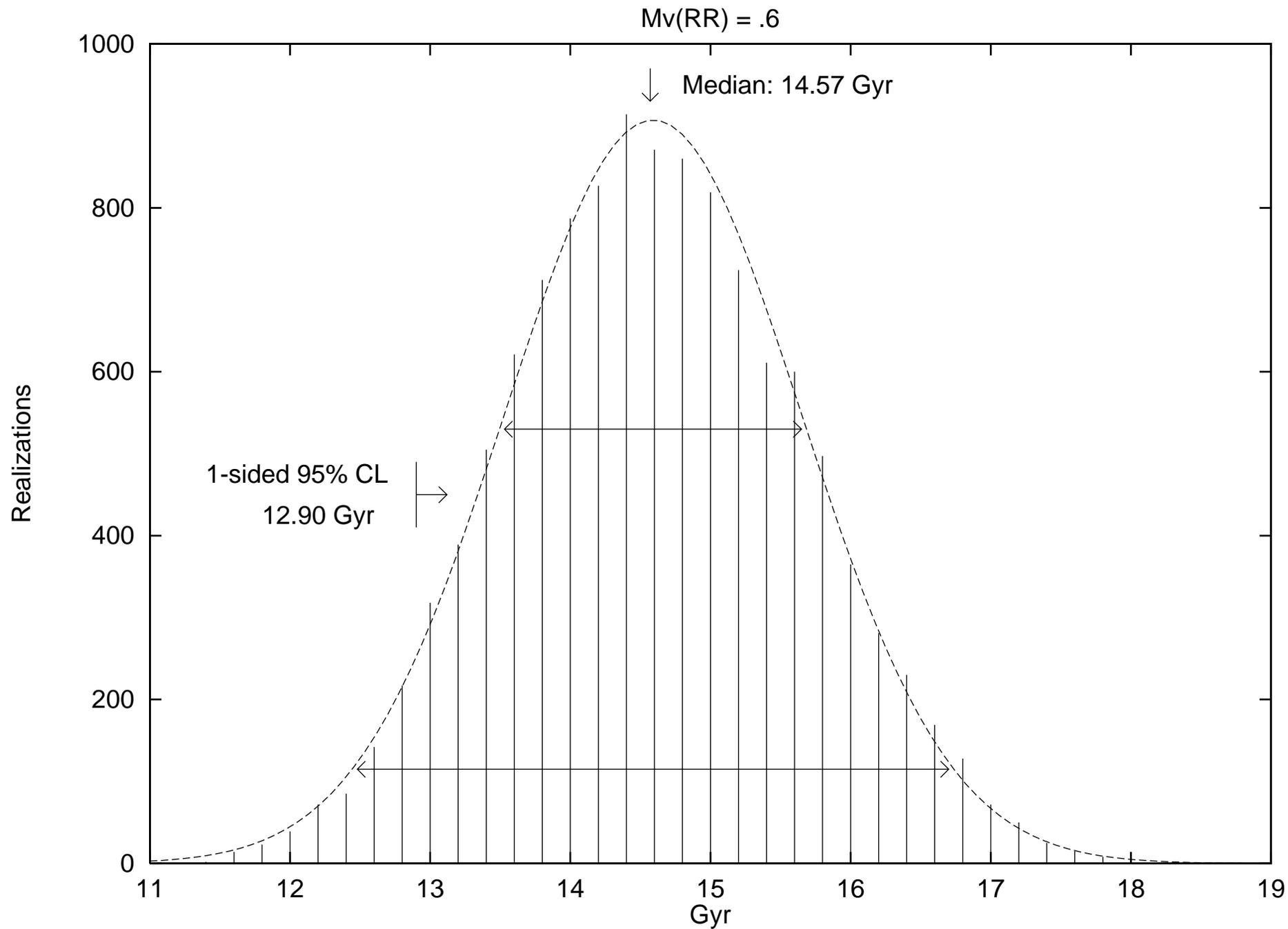
\end{document}